\documentclass[12pt]{article}

\usepackage[super,sort&compress,comma]{natbib} 

\usepackage[left=2.5cm, right=2.5cm, top=3cm, bottom=3cm]{geometry}

\usepackage{graphicx} 

\usepackage[english]{babel}

\usepackage[T1]{fontenc}
\usepackage{hyperref}
\usepackage[utf8]{inputenc}
\usepackage{amsmath}

\title{Preparation of graphene bilayers on platinum by \newline sequential chemical vapour deposition}
\author{
\noindent
Johannes Halle,$^{\ast}$ Alexander Mehler, Nicolas Néel, Jörg Kröger \\
\small{Institut für Physik, Technische Universität Ilmenau, D-98693 Ilmenau, Germany} \\ 
\small{E-mail: johannes.halle@tu-ilmenau.de}
}
\date{}

\begin{document}

\maketitle

\begin{abstract}
A cheap and flexible method is introduced that enables the epitaxial growth of bilayer graphene on Pt(111) by sequential chemical vapour deposition.
Extended regions of two stacked graphene sheets are obtained by, first, the thermal decomposition of ethylene and the subsequent formation of graphene.
In the second step, a sufficiently thick Pt film buries the first graphene layer and acts as a platform for the fabrication of the second graphene layer in the third step.
A final annealing process then leads to the diffusion of the first graphene sheet to the surface until the bilayer stacking with the second sheet is accomplished.
Scanning tunnelling microscopy unravels the successful growth of bilayer graphene and elucidates the origin of moiré patterns.
\end{abstract}

\section{Introduction}

Along with the ongoing interest in graphene and its derivatives comes the demand for reliable and scalable preparation methods for the eventual application in devices. 
For graphene-based layered systems such as bilayer graphene (BLG), whose properties depend strongly on stacking and interlayer twist angles, \cite{rpp_76_056503,natmater_12_887} the epitaxial growth directly on the sample yielded high structural quality. 

SiC substrates \cite{carbon_3_53,ss_48_463,jpd_43_374009} and transition metal surfaces \cite{jpcm_9_1,ss_603_1841} have widely been used for the preparation of monolayer graphene (MLG) as well as multilayer stackings.
For the fabrication of closed monolayers on metals, the thermal decomposition of hydrocarbons assisted by the catalytically active surface is typically performed as a chemical vapour deposition (CVD). \cite{njp_11_023006} 
The growth is self-limiting to one monolayer and, thus, offers precise thickness control. 
Additionally, this technique yields high-quality sheets with only a few rotational domains depending on substrate and growth temperature \cite{ss_603_1841,nl_8_565,apl_98_141903} and may readily be applied in industrial production. 
However, due to the self-limitation, its use for the preparation of multilayers is not straightforward. 
Instead, the growth of MLG and multilayer graphene on metals benefits from an alternative approach, namely the segregation of C at elevated temperatures.
To this end, the bulk metal is artificially enriched with C, either via doping from the gas phase of hydrocarbons \cite{jpcm_9_1,ss_603_1841,natmater_7_406,nl_9_30,pccp_12_5053} or solid-state diffusion. \cite{acsnano_4_1026,acsnano_5_1522,matdes_104_284,tsf_648_120} 
Alternatively, high-quality multilayer graphene on metals may also be achieved by intercalation of C under MLG, which requires sources of atomic C\@. \cite{acsnano_5_2298,nl_17_3105,small_14_1703701}

A preparation protocol that enables the growth of homogeneous as well as heterogeneous stackings of two-dimensional materials in a controlled surface science approach would be highly desirable.
Ideally, a CVD-based layer-by-layer preparation would combine the synthesis of large-scale graphene sheets with high structural quality and with precise thickness control owing to the self-limitation to one monolayer per growth cycle.

Here, we present a sequential CVD method that achieves the high-quality preparation of extended BLG regions on Pt(111)\@.
In a four-step process, MLG is first prepared on Pt(111) via thermal decomposition of C$_2$H$_4$, followed by the deposition of a thick Pt film and subsequent fabrication of a second graphene layer. 
During the postannealing in the last step, the diffusion of buried MLG to the surface is completed and leads to the formation of extended BLG domains. 
This last step can also be viewed as the intercalation of the deposited thick metal layer.
Various materials were previously demonstrated to intercalate graphene, e.\,g., H, \cite{prl_103_246804} Au, \cite{njp_12_033014,acsnano_8_3735} Co, \cite{prb_85_205434,prb_87_041403} Fe, \cite{acsnano_6_151} Ni, \cite{acsnano_6_151,prb_87_035420} Ag, \cite{prb_88_235430} Eu, \cite{nl_13_5013} Cs, \cite{nl_13_5013,natcommun_4_2772} as well as Li. \cite{jpcc_120_5067}
Therefore, Pt is very likely to intercalate, too.
 A method akin to the one suggested here has recently been applied to prepare MLG flakes on Au(111)\@. \cite{acsnano_8_3735} 
Since Au does not offer the convenient CVD synthesis, MLG flakes grown by CVD on Ir(111) were subsequently intercalated by several monolayers of Au to achieve graphene flakes on top of a Au(111) surface.
All prepared graphene layers are thoroughly characterised by atomically resolved scanning tunnelling microscopy (STM)\@.

\section{Results and discussion}

While in principle the presented preparation protocol should be applicable to Ir, Ru, Ni, \cite{jpcm_9_1,ss_603_1841} single-crystalline (111) surfaces of Pt have been chosen in this work since on this substrate the segregation after a CVD process is typically insufficient to already produce BLG.
Moreover, graphene on Pt(111) has the appealing property of the lowest graphene--metal coupling. \cite{prb_78_073401}

\begin{figure*}[thb]
\begin{center}
\includegraphics[width=\columnwidth]{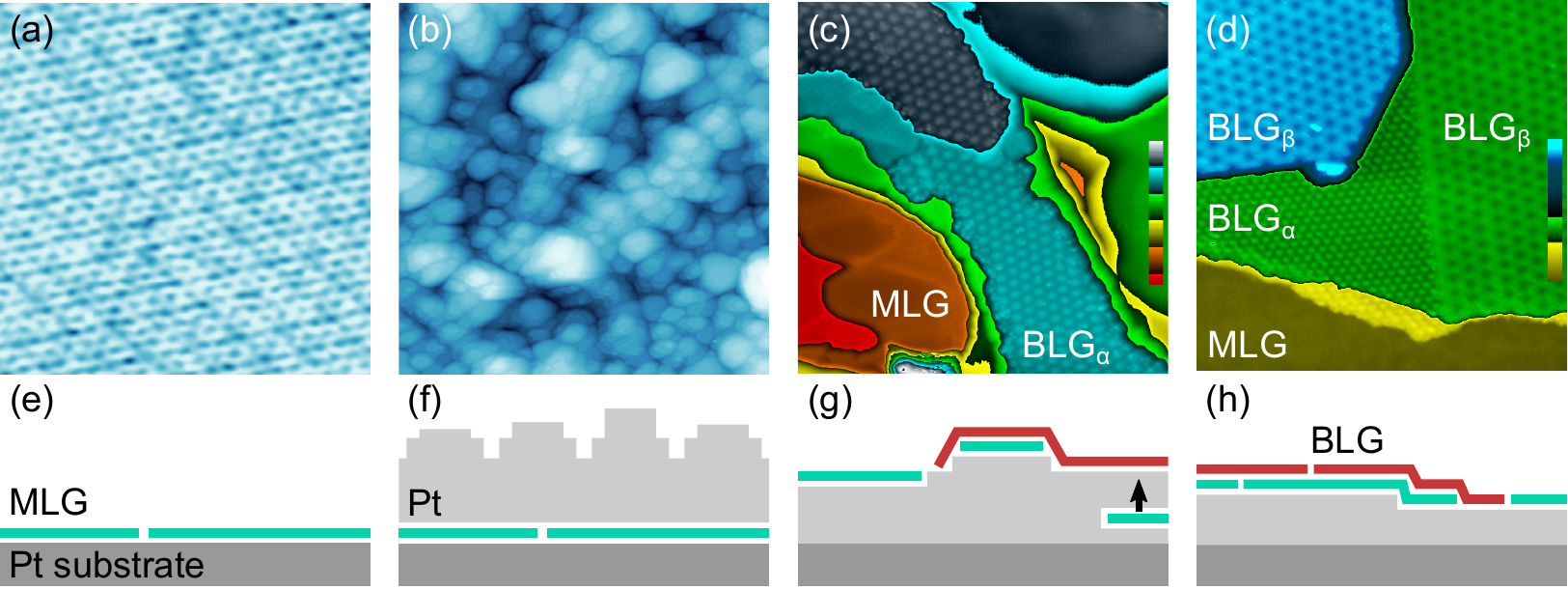}
\end{center}
\caption{
Steps of the BLG preparation protocol.
(a) Atomically resolved STM image of MLG on Pt(111) (bias voltage: $V=74\,\text{mV}$, tunnelling current: $I=100\,\text{pA}$, size: $5\times 5\,\text{nm}^2$)\@.
(b) STM image of Pt film deposited on MLG/Pt(111) ($1\,\text{V}$, $80\,\text{pA}$, $120\times 120\,\text{nm}^2$)\@.
(c) STM image of MLG and BLG regions after the second CVD step ($1\,\text{V}$, $80\,\text{pA}$, $40\times 40\,\text{nm}^2$)\@.
(d) STM image of MLG and BLG after annealing the sample of (c) ($1\,\text{V}$, $90\,\text{pA}$, $60\times 60\,\text{nm}^2$)\@.
(e)--(h) Illustration (cross-section) of the preparation steps.
}
\label{fig1}
\end{figure*}

Figure \ref{fig1} illustrates the preparation process by STM images (Fig.\,\ref{fig1}a--d) and sketches (Fig.\,\ref{fig1}e--h)\@.
First, MLG results from epitaxial growth of thermally decomposed C$_2$H$_4$ (Fig.\,\ref{fig1}a,e)\@.
In most regions, moiré patterns due to the lattice mismatch between graphene and Pt(111) were absent in STM images, which hints at large rotation angles of graphene with respect to Pt(111)\@. \cite{apl_98_033101,ss_606_1643}
Occasionally, moiré superstructures were observed and indicated smaller rotation angles.\cite{acsnano_5_5627} 
Second, approximately $45$ atomic layers of Pt were deposited on MLG-covered Pt(111).
STM images show resulting Pt clusters (Fig.\,\ref{fig1}b,f)\@.
In the third step, the second layer of graphene was grown by thermal decomposition of C$_2$H$_4$ on the Pt film.
The required annealing of the sample at $1000\,\text{K}$ led to the incomplete diffusion of the first graphene layer to the surface giving rise to large MLG and many small BLG regions (Fig.\,\ref{fig1}c,g)\@.
Further evidence that the buried graphene sheet forms the bottom layer of BLG in this preparation method is provided in the SI. 
The intercalation of Pt under MLG due to the high temperature during the CVD possibly breaks the MLG apart. 
Indeed, domain boundaries of MLG assist the intercalation process \cite{ss_617_81,natcommun_4_2772} and may likewise facilitate the rupture of MLG\@. \cite{jacs_134_5662}
Individual patches can then diffuse separately to the surface (Fig.\,\ref{fig1}g) as observed during the preparation of graphene nanoflakes on Au(111)/Ir(111). \cite{acsnano_8_3735}
Additionally, individual C atoms may detach from the boundary of buried graphene layers, segregate and contribute to the graphene growth at the surface. \cite{acsnano_4_1026,acsnano_5_1522,matdes_104_284,tsf_648_120}
The third step clearly demonstrates that a sufficiently thick Pt film is crucial to avoid completion of the Pt intercalation before the actual growth of the second graphene layer can start.
Subsequent annealing of the sample yielded flat and extended MLG and BLG regions (Fig.\,\ref{fig1}d,h)\@.
Second-layer graphene flakes hybridise to form extended BLG regions, while buried graphene patches continue their diffusion to the surface and may either be incorporated in the bottom graphene layer or create additional BLG or even trilayer graphene (TLG) regions.

Evidence for the proposed growth is given by the analysis of spatial periods ($\delta$) and the orientations of the moiré pattern ($\varphi$) and of the observed graphene lattice ($\vartheta$) with respect to Pt(111)\@.
The details of the analysis are summarised in the SI.
Moiré domains that do not match the expected characteristics of MLG are attributed to BLG.

\begin{figure*}[!t]
\begin{center}
\includegraphics[width=\columnwidth]{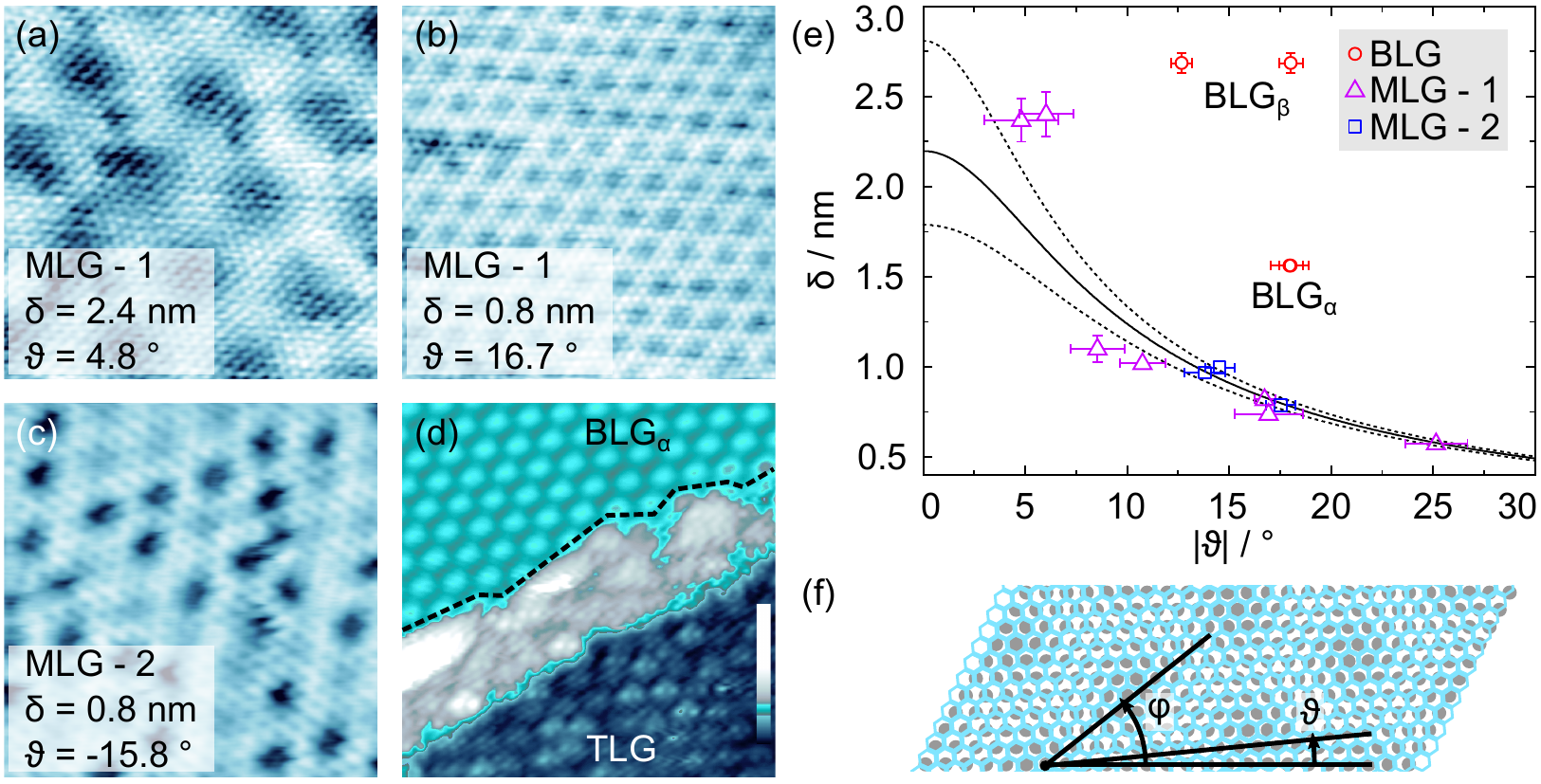}
\end{center}
\caption{
Overview and analysis of observed graphene domains on Pt(111) with indicated moiré spatial period ($\delta$) and graphene orientation ($\vartheta$)\@.
STM images of MLG domains with (a) $\delta=2.4\,\text{nm}$, $\vartheta=4.8^\circ$ ($100\,\text{mV}$, $100\,\text{pA}$, $8\times 8\,\text{nm}^2$) and (b) $\delta=0.8\,\text{nm}$, $\vartheta=16.7^\circ$ ($39\,\text{mV}$, $9\,\text{nA}$, $8\times 8\,\text{nm}^2$)\@.
(c) MLG domain after the final annealing step with $\delta=0.8\,\text{nm}$, $\vartheta=-15.8^\circ$ ($50\,\text{mV}$, $100\,\text{nA}$, $5\times 5\,\text{nm}^2$)\@.
The dark depressions are attributed to C atoms or vacancies in the Pt surface.
(d) Adjacent BLG and TLG domains ($400\,\text{mV}$, $100\,\text{pA}$, $20\times 20\,\text{nm}^2$)\@.
TLG crosses a substrate step edge and ends approximately at the boundary marked by the dashed line.
(e) Summary of $\delta$, $\vartheta$ for all observed moiré structures in MLG and BLG\@.
The solid line depicts the expected variation of $\delta$ with $\vartheta$ for unstrained MLG\@.
The upper (lower) dashed curve depicts $\delta$ as a function of $\vartheta$ for MLG with lattice constant increased (decreased) by $2\,\%$\@.
Open triangles (squares) are data for MLG after the first (second) CVD process, denoted as MLG-1 (MLG-2)\@.
Circles represent data for two BLG domains, BLG$_\alpha$ and BLG$_\beta$ (see text)\@.
(f) Solid lines mark crystallographic directions of Pt(111) (dots), graphene (circles) and the resulting moiré superlattice.
The orientations of graphene ($\vartheta$) and moiré lattice ($\varphi$) with respect to Pt(111) are indicated.
}
\label{fig2}
\end{figure*}

Figure \ref{fig2} summarises this analysis.
MLG after the first step of the preparation protocol (Fig.\,\ref{fig2}a,b) occasionally exhibits moiré patterns with, e.g., $\delta=2.37\,\text{nm}$, $\vartheta=4.8^\circ$ (Fig.\,\ref{fig2}a) and $\delta=0.82\,\text{nm}$, $\vartheta=16.7^\circ$ (Fig.\,\ref{fig2}b)\@.
After the complete preparation moiré patterns of MLG with $\delta=0.79\,\text{nm}$, $\vartheta=-15.8^\circ$ (Fig.\,\ref{fig2}c) are observed.
The dark depressions, which most clearly appear in the STM image of Fig.\,\ref{fig2}c, may be due to C atoms in the Pt surface \cite{tsf_648_120} or vacancies. \cite{prl_105_216102}
Figure \ref{fig2}d shows an STM image where BLG and TLG are present as adjacent regions.
The dashed line marks the boundary between the different domains.
All moiré spatial periods and orientations (symbols) together with the expected trends for MLG (solid and dashed lines) are summarised in Fig.\,\ref{fig2}e.
MLG data (triangles, squares) are reasonably well captured by the expected variation of $\delta$ with $\vartheta$ for unstrained MLG (Eq.\,S2). 
The dashed lines depict $\delta$ as a function of $\vartheta$ for a stretched (upper curve) and compressed (lower curve) C lattice, where the lattice constant deviates by $2\,\%$ from the unstrained case.
Deviations up to $7\,\%$ were reported for stable configurations of MLG on Pt(111)\@. \cite{acsnano_5_5627}
Figure \ref{fig2}f defines the angles enclosed by crystallographic directions of the moiré superlattice and graphene with respect to Pt(111)\@.

\begin{figure*}[!t]
\begin{center}
\includegraphics[width=\columnwidth]{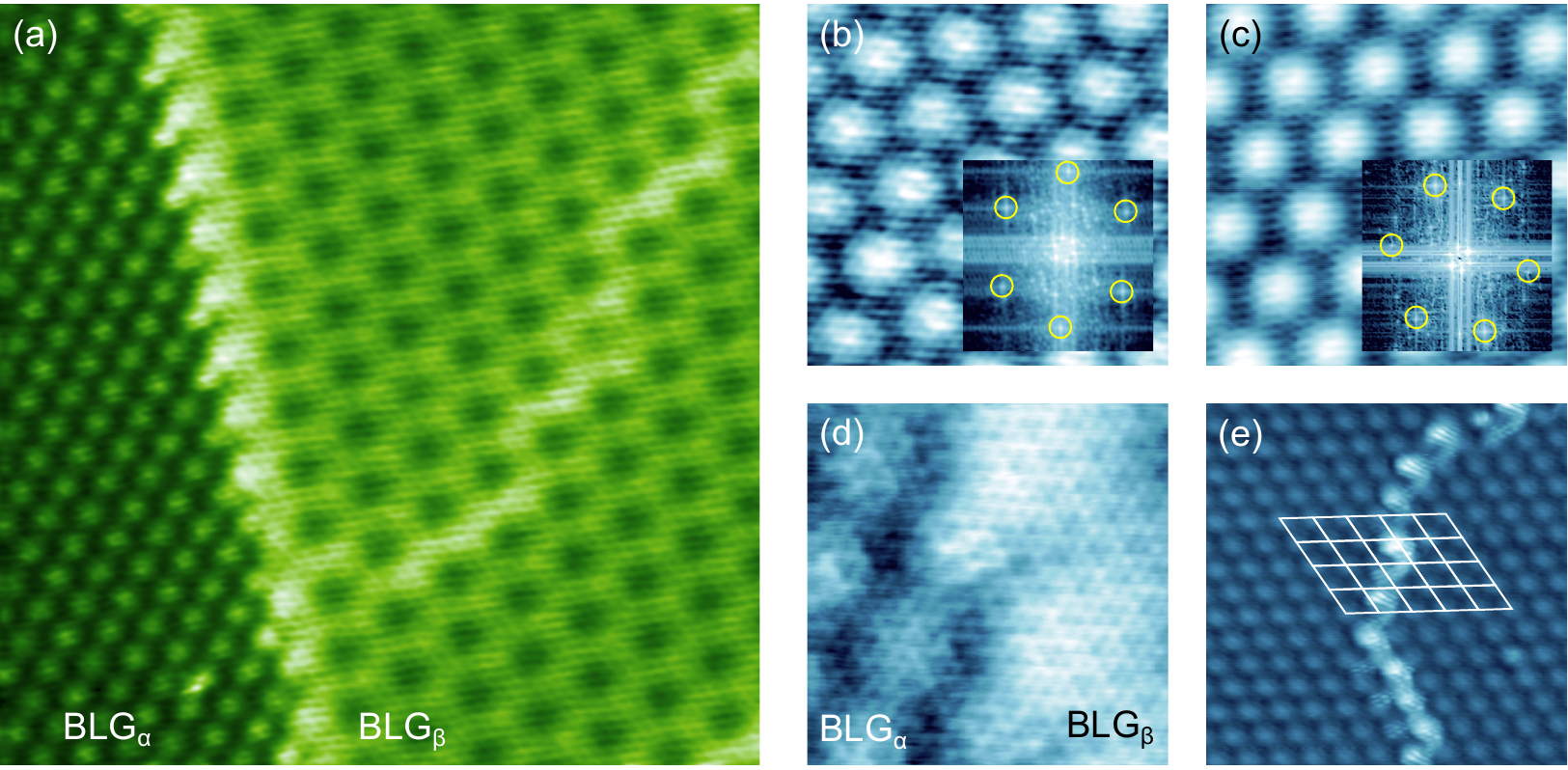}
\end{center}
\caption{
Graphene bilayer domains. 
(a) Overview STM image showing adjacent BLG$_\alpha$ and BLG$_\beta$ domains ($70\,\text{mV}$, $100\,\text{pA}$, $30\times 30\,\text{nm}^2$)\@.
(b), (c) STM images ($50\,\text{mV}$, $7\times 7\,\text{nm}^2$) of BLG$_\alpha$ subdomains with identical moiré characteristics ($\delta=1.6\,\text{nm}$, $\varphi=-28.8^\circ$) and different graphene orientations, (b) $\vartheta=18.0^\circ$ and (c) $\vartheta=-18.0^\circ$.
Tunnelling currents are $100\,\text{nA}$ (b) and $10\,\text{nA}$ (c)\@.
Insets: Fourier transforms of STM images with circles indicating the graphene lattice orientation.
(d) Atomically resolved STM image of the transition from the BLG$_\alpha$ ($\delta=1.6\,\text{nm}$, $\vartheta=18.0^\circ$, left) to the BLG$_\beta$ moiré superlattice ($\delta=2.7\,\text{nm}$, $\vartheta=18.0^\circ$, right) ($70\,\text{mV}$, $100\,\text{pA}$, $5\times 5\,\text{nm}^2$)\@.
(e) STM image of BLG$_\alpha$ with a line defect in the top graphene layer ($1\,\text{V}$, $80\,\text{pA}$, $20\times 20\,\text{nm}^2$)\@.
The superimposed lattice shows that the moiré pattern is continuous across the line defect.
}
\label{fig3}
\end{figure*}

Some data, however, deviate significantly from the expected variation; that is, in an MLG model the observed moiré spatial period is not compatible with the observable $\vartheta$.
The corresponding domains are only found after the second CVD process and are therefore assigned to BLG\@. The gallery of STM images in Fig.\,\ref{fig3} further corroborates this assignment.
Figure \ref{fig3}a shows an extended BLG region that contains both moiré patterns, BLG$_\alpha$ ($\delta=1.6\,\text{nm}$, $\varphi=-28.8^\circ$) and BLG$_\beta$ ($\delta=2.7\,\text{nm}$, $\varphi=0.8^\circ$)\@. 
BLG$_\alpha$ and BLG$_\beta$ each exhibit two subdomains with identical $\delta$ and $\varphi$ but different $\vartheta$. 
Examplarily, this is demonstrated in Fig.\,\ref{fig3}b,c for the BLG$_\alpha$ domain with atomic resolution. 
The Fourier transforms (insets to Fig.\,\ref{fig3}b,c) reveal that the graphene lattices of the subdomains are rotated by $36^\circ$ with respect to each other, while the moiré pattern is unaffected. 
These observations unambiguously demonstrate the presence of two graphene sheets in BLG$_\alpha$ and BLG$_\beta$ domains.

Remarkably, an atomically resolved close-up view of the domain boundary between BLG$_\alpha$ and BLG$_\beta$ (Fig.\,\ref{fig3}d) reveals a unique graphene lattice covering both domains. Here, subdomains of BLG$_\alpha$ and BLG$_\beta$ with $\vartheta=18^\circ$ are connected.
The uniformly oriented top layer of graphene demonstrates that the different moiré patterns must be due to rotational domains of the bottom graphene sheet.
The STM image depicted in Fig.\,\ref{fig3}e corroborates this important aspect. 
It shows a BLG$_\alpha$ area with a line defect in the top graphene sheet. 
The moiré pattern ($\delta=1.6\,\text{nm}$, $\varphi=-28.8^\circ$), however, remains unaffected across the defect as evidenced by the superimposed lattice. 
Therefore, the observed moiré superstructure is due to the graphene/Pt(111) interface and, intriguingly, the upper graphene layer does not contribute to the moiré pattern.

Previously, moiré lattices on BLG/Ir(111) were attributed to the graphene/graphene interface. \cite{small_14_1703701} 
The apparent contradiction with findings reported here is resolved when \textit{both} the graphene/substrate \textit{and} graphene/graphene twist angles are considered.
For both these interfaces, low twist angles ($<10^\circ$) give rise to an elevated hybridization, while large twist angles ($>10^\circ$) decouple the layers effectively. \cite{prl_100_125504,prl_106_126802,prb_83_125428,small_14_1703701}
In BLG$_\alpha$ and BLG$_\beta$ domains the bottom graphene layer is rotated by only $4.6^\circ$ and $0.1^\circ$ with respect to the Pt(111) surface, respectively (Eq.\,S4), while all graphene/graphene twist angles exceed $10^\circ$\@.
In contrast, most of the reported domains of BLG/Ir(111) exhibit large twist angles at both interfaces. \cite{small_14_1703701}
Consequently, the elevated (low) graphene-Pt (graphene-Ir) hybridization favours the observation of the moiré pattern due to the graphene/Pt (graphene/graphene) interface.
The twist angles at both interfaces and the resulting interlayer coupling therefore determine which moiré superstructure is visible in STM images.

\section{Conclusions}

In conclusion, a preparation protocol for the preparation of extended BLG on Pt(111) has been introduced.
The applicability of the method to other transition metal substrates and homogeneous as well as heterogeneous multistackings of two-dimensional materials may be anticipated.
The presented atomically resolved STM data unveil that the moiré superlattices observed in BLG regions result from the graphene/Pt(111) interface. 

\section{Experimental methods}

Pt(111) surfaces were prepared by Ar$^+$ bombardment and annealing at $1200\,\text{K}$ in O$_2$ atmosphere ($4\cdot 10^{-5}\,\text{Pa}$)\@. 
C$_2$H$_4$ at a partial pressure of $4\cdot 10^{-5}\,\text{Pa}$ was used as a molecular precursor in the CVD process. 
After MLG growth via thermal decomposition of C$_2$H$_4$ at $1200\,\text{K}$, Pt was evaporated from a hot filament at a rate of $\approx 1\,\text{ML}/\text{min}$ (ML: monolayer), which was calibrated from STM images of several independent depositions on Au(100) and MLG/Pt(111)\@. 
The preparation of the second graphene layer was performed by annealing the sample at $1000\,\text{K}$ for $3\,\text{min}$, directly followed by exposure to C$_2$H$_4$ for $4\,\text{min}$. 
The sample was kept at $1000\,\text{K}$ for $5\,\text{min}$ before cooling to room temperature. 
Additional annealing cycles were performed at $1200\,\text{K}$\@.
The experiments were performed in ultrahigh vacuum (base pressure $1\cdot 10^{-8}\,\text{Pa}$) with a low-temperature STM operated at $6\,\text{K}$\@. 
Tips were cut from Au wire and cleaned in situ by annealing and field emission. 
STM images were recorded in the constant-current mode with the bias voltage applied to the sample.
Topographic data were processed using the WSxM software. \cite{rsi_78_013705}

\section*{Conflicts of interest}
There are no conflicts to declare.

\section*{Acknowledgements}
Financial support by the Deutsche Forschungsgemeinschaft through Grants No.\,KR\,$2912/10-1$ and KR\,$2912/12-1$  is acknowledged.


\end{document}


\maketitle

\section{Intercalation of deposited Pt film}

\begin{figure}[htb]
\begin{center}
\includegraphics[width=0.9\linewidth]{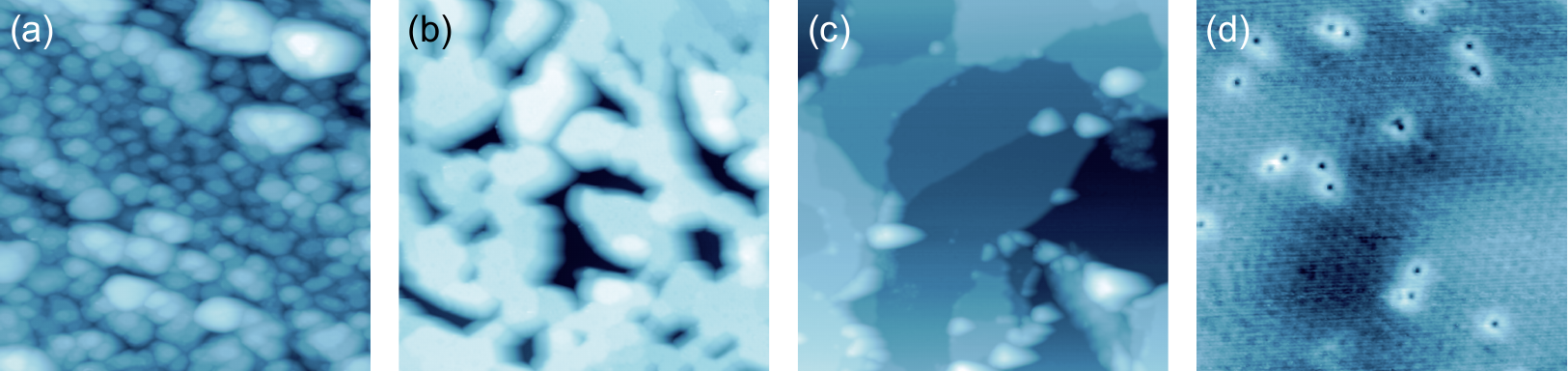}
\end{center}
\caption{
Pt intercalation of MLG-covered Pt(111)\@.
(a) STM image of Pt film deposited on MLG/Pt(111) ($1\,\text{V}$, $80\,\text{pA}$, $120\times 120\,\text{nm}^2$)\@.
(b) STM image of the deposited Pt smoothed by annealing at $720\,\text{K}$ ($1.5\,\text{V}$, $80\,\text{pA}$, $150\times 150\,\text{nm}^2$)\@.
(c) STM image of MLG/Pt(111) after complete intercalation of the deposited Pt at $1200\,\text{K}$ ($1.5\,\text{V}$, $80\,\text{pA}$, $100\times 100\,\text{nm}^2$)\@.
(d) MLG moiré pattern observed on the sample shown in (c) ($0.1\,\text{V}$, $100\,\text{pA}$, $20\times 20\,\text{nm}^2$)\@.
}
\label{figS1}
\end{figure}

In order to further evidence that the buried monolayer graphene (MLG) forms the bottom layer in bilayer graphene (BLG), Fig.~\ref{figS1} demonstrates the entire intercalation of the thick Pt film through the buried MLG\@. 
Upon annealing the Pt-covered MLG/Pt(111) sample (Fig.~\ref{figS1}a) at $1200\,\text{K}$, the deposited Pt is flattening and intercalating under MLG (Fig.~\ref{figS1}b,c)\@. 
The segregated MLG exhbits the characteristic moiré pattern (Fig.~\ref{figS1}d)\@. 

\section{Analysis of graphene moiré patterns}
\subsection*{Moiré patterns of monolayer graphene on Pt(111)}

The moiré pattern is generated by the superposition of Pt(111) and the monolayer graphene (MLG) lattices. 
In direct space, $a_{\text{Pt}}$, $a_\text{C}$, $\delta$ denote the spatial periods of, respectively, the Pt(111) surface, the graphene lattice, the moiré superstructure.
In reciprocal space, the corresponding lattice vectors are $\mathbf{k}_{\text{Pt}}$, $\mathbf{k}_\text{C}$, $\mathbf{k}_\text{m}$ with magnitudes $k_{\text{Pt}}=\vert\mathbf{k}_{\text{Pt}}\vert=1/a_{\text{Pt}}$, $k_\text{C}=\vert\mathbf{k}_\text{C}\vert=1/a_\text{C}$, $k_\text{m}=\vert\mathbf{k}_\text{m}\vert=1/\delta$.
The moiré superstructure represents a spatial beating pattern, which may be expressed by \cite{njp_10_043033}
\begin{equation}
\mathbf{k}_\text{m}=\mathbf{k}_\text{C}-\mathbf{k}_{\text{Pt}}.
\label{eqS1}
\end{equation}
with $\vartheta=\angle(\mathbf{k}_{\text{Pt}},\mathbf{k}_\text{C})$ and $\varphi=\angle(\mathbf{k}_{\text{Pt}},\mathbf{k}_\text{m})$ defined as the smallest angles enclosed by the respective lattice orientations (Fig.~\ref{figS1}). 
Due to the hexagonal symmetry of the lattices, $\vartheta$ and $\varphi$ are constrained to the interval $[-30^\circ,30^\circ]$\@.
Spatial periods and orientations of the moiré pattern may readily be derived from the triangle formed by the three reciprocal lattice vectors (Fig.~\ref{figS1}a)\@. 
Using the law of cosines yields $\delta$ as a function of $\vartheta$,
\begin{equation}
\delta=\sqrt{\frac{1}{a_\text{C}^2}+\frac{1}{a_\text{Pt}^2}-\frac{2\cos\vartheta}{a_\text{C}\cdot a_\text{Pt}}}^{-1}.
\label{eqS2}
\end{equation}
Possible strain in the graphene sheet can be included by scaling the graphene lattice constant appropriately. 
Fig.~2e of the manuscript shows the resulting $\delta$-versus-$\vartheta$ curves. 
Since $\delta$ is an even function of $\vartheta$, the interval $\vartheta=[0^\circ,30^\circ]$ contains all possible moiré patterns.

\begin{figure}[thb]
\begin{center}
\includegraphics[width=0.9\linewidth]{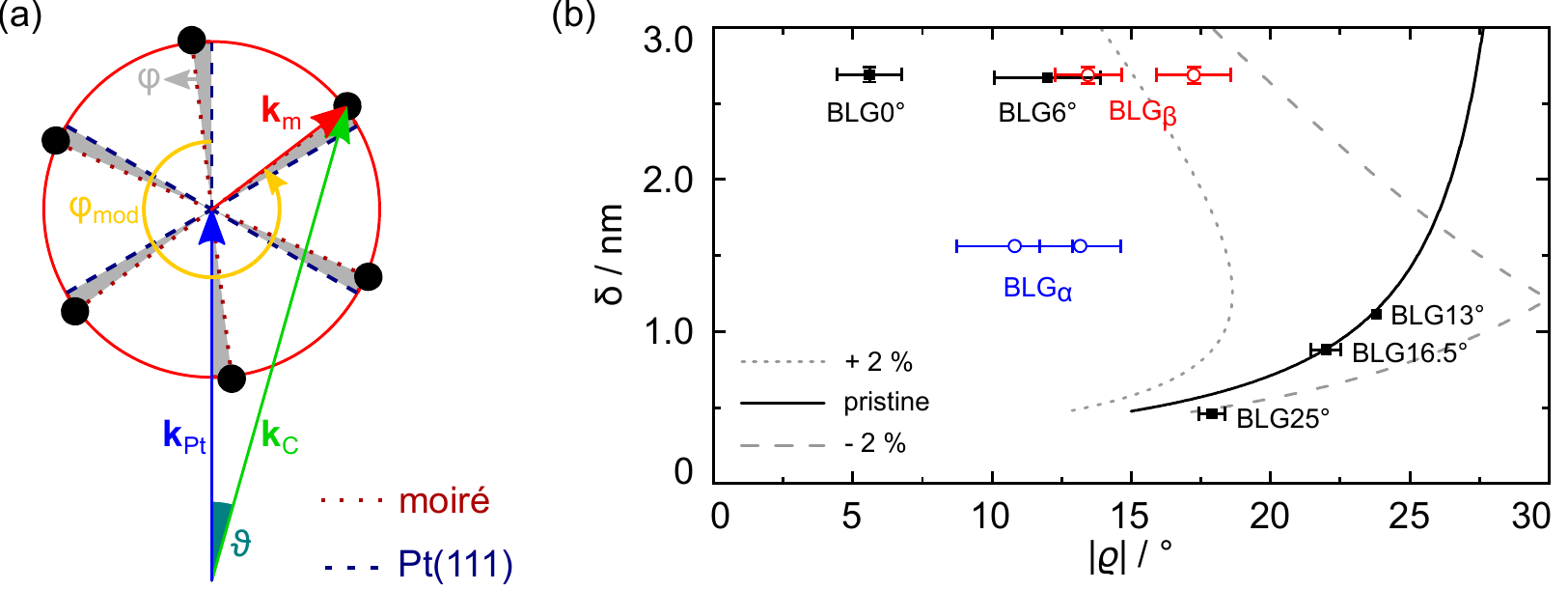}
\end{center}
\caption{
(a) Illustration of the moiré construction. 
The reciprocal lattice vectors of Pt(111) ($\mathbf{k}_{\text{Pt}}$), graphene ($\mathbf{k}_\text{C}$) and the moiré pattern ($\mathbf{k}_\text{m}$) form a triangle. 
The angle enclosed by crystallographic directions of Pt(111) and graphene is denoted as $\vartheta=\protect\angle(\mathbf{k}_{\text{Pt}},\mathbf{k}_\text{C})$, while $\varphi=\protect\angle(\mathbf{k}_{\text{Pt}},\mathbf{k}_\text{m})$ is the angle between crystallographic directions of Pt(111) and the moiré lattice.
Dashed and dotted lines represent symmetry directions of, respectively, Pt(111) and the moiré superstructure. 
Dots on the circle with radius $k_\text{m}=\vert\mathbf{k}_\text{m}\vert$ indicate equivalent moiré superstructures with $\varphi_{\text{mod}}=\varphi+n\cdot 60^\circ$ ($n=1,2,\ldots,5$), which reproduce the same observable moiré characteristics for varying $\mathbf{k}_\text{C}$.
(b) Twisted bilayer graphene (BLG) model. 
Lines depict the calculated moiré spatial period $\delta$ as a function of the angle $\varrho$ enclosed by crystallographic directions of the upper graphene lattice and the moiré superstructure.
The solid line shows the variation of $\delta$ with $\vert\varrho\vert$ for an unstrained lower graphene sheet, whereas dotted and dashed lines represent the situation for a graphene lattice constant that is increased and decreased by $2\,\%$, respectively.
Experimental data appear as circles (this work) and squares (BLG/Ir(111)) \cite{small_14_1703701}.
The angles $0^\circ\ldots\,25^\circ$ denote the twist angles between the bottom and upper graphene sheets.
}
\label{figS2}
\end{figure}

\subsection*{Characterization of the bottom graphene layer in bilayer graphene domains} 

An important conclusion of the findings presented in the main manuscript is the origin of the moiré pattern observed from bilayer graphene (BLG) domains.
The results evidence that the moiré superstructure is caused by the interface beween the bottom graphene layer and Pt(111)\@.
Using the moiré characteristics observed from BLG (BLG$_\alpha$, BLG$_\beta$ in Fig.~2e), the spatial period and orientation of the bottom graphene layer may be deduced as follows.
Given the spatial moiré period~$\delta$ and its orientation~$\varphi$ with respect to the Pt(111) lattice, six moiré orientations defined by $\varphi_{\text{mod}}$ in Fig.~\ref{figS1}a are indistinguishable in the experiments.
These orientations differ by $60^\circ$\@.
Free parameters are the unknown orientation of the bottom graphene layer, $\vartheta_\text{b}$, and its lattice constant, $a_\text{C}$, which may be affected by strain. 
The solutions with minimal strain are considered most plausible and can be obtained by calculating $a_\text{C}$ for every $\varphi_{\text{mod}}$, 
\begin{equation}
a_\text{C}=\sqrt{\frac{1}{a_{\text{Pt}}^2}+\frac{1}{\delta^2}+\frac{2\cos\varphi_{\text{mod}}}{a_{\text{Pt}}\cdot\delta}}^{-1}.
\label{eqS3}
\end{equation}
Subsequently, $\vartheta_\text{b}$ is calculated for the selected $a_\text{C}^{\text{min}}$ with minimal strain at the angle $\varphi_{\text{mod}}^{\text{min}}$ using
\begin{equation}
\delta\sin\vartheta_\text{b}=a_\text{C}^{\text{min}}\sin\varphi_{\text{mod}}^{\text{min}}.
\label{eqS4}
\end{equation}
Following this procedure, the bottom graphene sheets of the observed BLG domains are characterized by $\vartheta_\text{b}=4.6^\circ$ with $a_\text{C}=240\,\text{pm}$ ($-2.5\,\%$ strain) in the case of BLG$_\alpha$ and $\vartheta_\text{b}=0.1^\circ$ with $a_\text{C}=252\,\text{pm}$ ($2.1\,\%$ strain) in the BLG$_\beta$ domain. 
Consequently, the twist angles $\theta$ between adjacent graphene layers may now be obtained as $\theta=22.6^\circ$, $13.4^\circ$ for BLG$_\alpha$ and $\theta=12.6^\circ$, $18.1^\circ$ for BLG$_\beta$ subdomains.

\subsection*{Moiré patterns of twisted bilayer graphene}

The moiré pattern generated by twisted BLG can be calculated using Eq.~\ref{eqS2} with $a_{\text{Pt}}$ replaced by $a_\text{C}$. 
Thus, the spatial period of a moiré pattern resulting from two unstrained graphene lattices is described by
\begin{equation}
\delta=\frac{a_\text{C}}{\sqrt{2(1-\cos\theta)}}.
\label{eqS5}
\end{equation}

Assuming that the twist angle between bottom (b) and top (t) graphene sheets causes the observed moiré pattern, b-graphene may be characterized on the basis of this moiré pattern. 
To this end, t-graphene takes the role of the Pt lattice in Eqs.~\ref{eqS3}, \ref{eqS4} with known lattice constant $\text{a}_{\text{C,t}}$ and as a reference lattice for all angles. 
The lattice constant $a_{\text{C,b}}$ of the lower graphene is described by
\begin{equation}
a_{\text{C,b}}=\sqrt{\frac{1}{a_{\text{C,t}}^2}+\frac{1}{\delta^2}+\frac{2\cos\varrho_{\text{mod}}}{a_{\text{C,t}}\cdot\delta}}
\label{eqS6}
\end{equation}
for each of the six possible orientations defined by $\varrho_{\text{mod}}$, which describes the angle enclosed by the moiré superstructure and t-graphene.
The twist angle $\theta$ between the two graphene sheets is then given by 
\begin{equation}
\delta\sin\theta=a_{\text{C,b}}^{\text{min}}\sin\varrho_{\text{mod}}^{\text{min}}.
\label{eqS7}
\end{equation}

Fig.~\ref{figS1}b shows BLG data obtained in this work as well as from a recent study of BLG on Ir(111), where the observed moiré patterns were attributed to twisted graphene bilayers \cite{small_14_1703701}.
While data for larger twist angles ($>13^\circ$) follow the expected behavior for pristine BLG reasonably well, there is a notable deviation at smaller rotation angles. 
Indeed, the BLG0$^\circ$ moiré superstructure is virtually that of the underlying graphene/substrate interface. 
Density functional calculations \cite{small_14_1703701} showed that low twist angles give rise to an elevated graphene-metal coupling, which is consistent with the mechanism discussed in the main manuscript. 
In the twisted-BLG model, the BLG$_\alpha$ subdomains (Fig.~3c,d) would require comparably large strains of $3.5\,\%$ and $4.2\,\%$, but the BLG$_\beta$ subdomains would exhibit only $1.6\,\%$ and $2.3\,\%$ strain. 
However, the resulting moiré patterns of the subdomains would have to align in order to reproduce the experimental observations.
Therefore, the twisted-BLG model is not appropriate to explain the observed moiré structures.
